\documentclass[12pt,preprint]{aastex}

\shorttitle{Solar flare electron distribution functions}
\shortauthors{MB \& EPK}
\begin{document}
\title{ELECTRON DISTRIBUTION FUNCTIONS IN SOLAR FLARES FROM COMBINED X-RAY AND EUV OBSERVATIONS}
\author{M. Battaglia \altaffilmark{1}, E. P. Kontar \altaffilmark{2}}
\affil{Institute of 4D Technologies, School of Engineering, University of Applied Sciences and Arts Northwestern Switzerland, 5210 Windisch, Switzerland}
\affil{SUPA, School of Physics and Astronomy, University of Glasgow, G12 8QQ, UK}
\email{e-mail: marina.battaglia@fhnw.ch}
\begin{abstract}
Simultaneous solar flare observations with SDO and RHESSI provide spatially resolved information about hot plasma and energetic particles in flares. RHESSI allows the properties of both hot ($\gtrsim$ 8 MK) thermal plasma and nonthermal electron distributions to be inferred, while SDO/AIA is more sensitive to lower temperatures. We present and implement a new method to reconstruct electron distribution functions from SDO/AIA data. The combined analysis of RHESSI and AIA data allows the electron distribution function to be inferred over the broad energy range from ~0.1 keV up to a few tens of keV. The analysis of two well observed flares suggests that the distributions in general agree to within a factor of three when the RHESSI values are extrapolated into the intermediate range 1-3 keV, with AIA systematically predicting lower electron distributions. Possible instrumental and numerical effects, as well as potential physical origins for this discrepancy are discussed. 
The inferred electron distribution functions in general show one or two nearly Maxwellian components at energies below $\sim$ 15 keV and a non-thermal tail above. \\
\end{abstract}
\maketitle
\section{INTRODUCTION}
Solar flares accelerate large numbers of particles within seconds and maintain 
efficient acceleration over time-scales of minutes to tens of minutes. Hard X-ray (HXR) observations of solar flares provide information on the electron acceleration and transport in solar flares and are crucial to better constrain various flare models \citep[e.g.][as a recent review]{2011SSRv..159..107H}. Observations with RHESSI \citep{2002SoPh..210....3L} 
have significantly enriched our understanding of solar flare physics by providing accurate spatially resolved soft X-ray (SXR) and HXR spectra, which are used as the input to deduce the properties of the accelerated electrons \citep[e.g.][as a recent review]{2011SSRv..159..301K}. RHESSI also makes it possible to analyze individual source spectra (coronal sources, footpoints), by means of imaging spectroscopy \citep[e.g.][]{Em03,Ba06}. The technique was recently used by \citet{2013A&A...551A.135S} to infer the characteristic electron acceleration rates at the looptop and at the footpoints of a number of flares finding strong indication for particle trapping at the top of flaring loops. However, RHESSI detectors only provide X-ray observations above 3~keV, or even higher energies during large flares when attenuators are placed in front of the detectors to reduce dead-time which results in much reduced flux at energies below $\sim$ 6-12 keV \citep{Smith02}. While priceless to infer the properties of deka-keV electrons it leaves an observational gap for electrons below 3~keV. Due to the typical shape of flare spectra particles at these energies constitute the larger part of the total particle number in flares. It is important to extend the diagnostics to energies below 3 keV to answer questions such as whether the distribution at these energies is an iso-thermal Maxwellian. 
Unlike the higher X-ray energies, hot flaring plasma is normally studied using EUV and X-ray line emissions. Observations of the line intensities provide information on the temperature distribution of the solar plasma. With the recent launch of the Atmospheric Imaging Assembly (AIA) on board the Solar Dynamics Observatory \citep[SDO,][]{Le11} it is now possible to deduce differential emission measures (DEM) over a broad range of 
coronal temperatures with high spatial resolution. Complementing SXR and HXR observations with EUV observations allows the mean electron flux spectrum to be deduced over a wide range of energies from as low as 0.1 keV up to tens of keV.

In this paper we develop a method to infer the low energy part of the electron flux spectrum from DEMs and apply the method to spatially resolved AIA observations. We first demonstrate the method on a set of synthetic DEMs for which the electron distribution function is known and then apply it to flare observations. Using simultaneous X-ray and EUV observations we deduce, for the first time, 
the mean electron flux spectrum and estimate the local electron distribution function in various regions of the flaring atmosphere. The observations show that locally non-thermal electrons constitute a relatively 
modest fraction of the order of $10^{-2}$ of the total electron numbers. Different flaring regions demonstrate distinct electron distributions: footpoint regions
can be viewed consistent with a Maxwellian distribution and a non-thermal tail while an above-the-looptop source deviates from a Maxwellian in the range above 0.5 keV. 
\section{RECONSTRUCTION OF ELECTRON DISTRIBUTION FUNCTION FROM DEM(T)} \label{sec:theory}
Using line observations the properties of thermal plasma are traditionally characterized by the differential emission measure $\xi(T)$
($\mathrm{cm^{-3}K^{-1}}$). Assuming a locally Maxwellian distribution with local temperature $T(r)$ and density $n(r)$ at spatial coordinate r, $\xi(T)$ characterizes the amount of plasma at various plasma temperatures $T(r)$, so that the integral over $T$ gives the total emission measure $EM=\int \xi(T)dT$.
The Maxwellian distribution at temperature $T(r)$ is given as: 
\begin{equation} \label{eq:maxwell}
F(E,r)=\frac{2^{3/2}}{(\pi m_e)^{1/2}}\frac{n(r)E}{(k_BT(r))^{3/2}}\exp {(-E/k_BT(r))}, 
\end{equation}
where $E$ is the electron kinetic energy, $m_e$ is the electron mass, and $k_b$ is the Boltzmann constant.
This is related to the mean electron flux spectrum $\langle nVF\rangle$ in the emitting volume $V$ and the differential emission measure $\xi(T)$ following e.g. \citet{1988ApJ...331..554B}:
\begin{eqnarray} \label{eq:intoverdv}
\langle nVF\rangle &=&\int_Vn(r)F(E,r) dV \\
&=&\int_Tn(r)\frac{2^{3/2}}{(\pi m_e)^{1/2}}\frac{n(r)E}{(k_BT(r))^{3/2}}\exp {(-E/k_BT(r))}\frac{dV}{dT}dT 
\end{eqnarray}
where $n^2dV/dT=\xi(T)$ is the differential emission measure and thus:
\begin{equation} \label{eq:meaneflux}
\langle nVF\rangle =\frac{2^{3/2}E}{(\pi m_e)^{1/2}}\int_0^{\infty}\frac{\xi(T)}{(k_BT)^{3/2}}\exp{(-E/k_BT)}\mathrm{dT}.
\end{equation}
Therefore, knowing the differential emission measure one can compute the electron flux spectrum in the emitting volume ($\mathrm{electrons~keV^{-1}s^{-1}cm^{-2}}$).
Although Equation (\ref{eq:meaneflux}), which is an equivalent of the Laplace transform of a function $f(t)$
\begin{equation} \label{eq:basicf}
F(s)=\int_0^\infty \exp(-st)f(t)\mathrm{dt},
\end{equation}
is formally a straightforward integration over temperature,
the numerical integration could be rather challenging due to the exponential kernel \citep[e.g.][]{2006SoPh..237...61P}.
Following \citet{Ros08}, we rewrite the Laplace transform (Equation (\ref {eq:basicf})) via 
the convolution integral, which will allow efficient numerical computations of $\langle nVF \rangle$ via $\xi(T)$ and vice versa.
Using the change of variables $s=\exp(y)$ and $t=\exp(-x)$, let us rewrite Equation (\ref{eq:basicf}) in the following form: 
\begin{equation} \label{eq:theory}
F(e^y)=\int_{-\infty}^{\infty}K(y-x)h(x)\mathrm{d}x
\end{equation}
where $K(y-x)=\exp(y-x)\exp[-\exp(y-x)]$ and $h(x)=\phi(e^{-x})$ with $\phi(t)=\int_0^t f(t')\mathrm{dt'}$.
Equation (\ref{eq:meaneflux}) can be similarly brought into the form of Equation (\ref{eq:basicf}) 
using the variable change
\begin{equation}
t=1/T \, ; \frac{dt}{dT}=-\frac{1}{T^2} \, ; dT=-\frac{1}{t^2}dt
\end{equation}
which results in  
\begin{equation} \label{eq:final}
\langle nVF\rangle =\frac{2^{3/2}E}{(\pi m_e)^{1/2}k_B^{3/2}}\int_{0}^{\infty}\frac{\xi(T(t))}{t^{1/2}}\exp{(-Et/k_B)}\mathrm{d}t,
\end{equation}
so that $f(t)=\frac{\xi(T(t))}{t^{1/2}}$ and $\exp(-st)=\exp(-Et/k_B)$ in Equation (\ref{eq:basicf}), which is then brought into the form of Equation (\ref{eq:theory}) and solved.
\subsection{APPLICATION ON SYNTHETIC DEM}
\begin{figure}[h!]
\includegraphics[height=14cm]{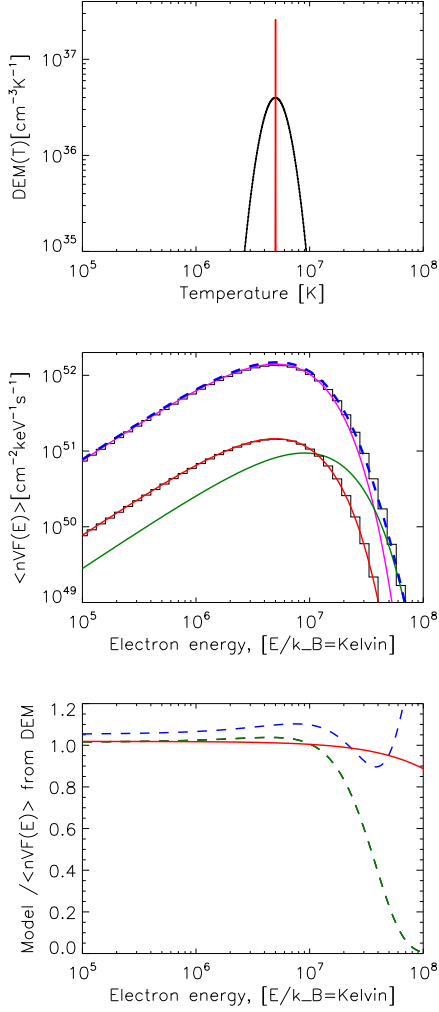}
\caption {Top: Synthetic DEM $\mathrm{(cm^{-3}K^{-1})}$ as function of $T$ for peak temperature 5 MK and two different widths (red: $\delta$-function, black: $\sigma=0.1$, compare Equation \ref{eq:gaussdem}). Middle: Reconstructed mean electron flux spectrum from DEM (black lines). The total EM of the DEM with width  $\sigma=0.1$ was chosen one order of magnitude larger then the $\delta$-function to give clearly distinguishable electron spectra. The red line gives a Maxwellian distribution at temperature 5 MK. The purple and green lines are Maxwellian distributions at 5 MK and 9 MK, the dashed blue line is the sum of these two Maxwellians.  Bottom: Model flux divided by flux from DEM in the case of a delta-function DEM (red solid line), and in the case of DEM of width $\sigma=0.1$ relative to a single temperature Maxwellian (dashed green) and two Maxwellians at 5 MK and 9 MK (dashed blue line). }
\label{fig:synDEM}
\end{figure}

We illustrate the method using two synthetic DEMs. The first is a single temperature DEM, i.e. a $\delta$-function in temperature space (Figure \ref{fig:synDEM}, top) at temperature $T_0=5$ MK. 
The mean electron flux spectrum corresponding to this DEM is calculated using Equations (\ref{eq:meaneflux} - \ref{eq:final}). The result is shown in the middle panel of Figure \ref{fig:synDEM}. This is compared to the Maxwellian distribution as defined in Equation (\ref{eq:maxwell}).
From Equation (\ref{eq:intoverdv}) one finds the mean electron flux spectrum for a uniform distribution over the whole volume as:
\begin{equation}
\langle nVF\rangle =n_e^2V\left(\frac{2}{k_BT}\right)^{3/2}\frac{E}{(\pi m_e)^{1/2}}\exp {(-E/k_BT)},
 \label {eq:totmaxwell}
\end{equation}
where $n_e^2V=EM$ is the total emission measure. As Figure \ref{fig:synDEM} indicates, there is a very close agreement (better than 2 \% except for temperatures above $\sim 2\times 10^7$ K, where numerical effects become noticeable) between the analytical solution (Equation (\ref{eq:totmaxwell})) and the mean electron flux spectrum found by integrating the DEM (Equation (\ref{eq:final})), validating the method.
The second synthetic DEM for which the method is demonstrated is a Gaussian DEM-function in $\log T$: 
\begin{equation} \label{eq:gaussdem}
\xi(T)=\frac{\xi_0}{\sqrt{2\pi}\sigma^2}\exp(-(\log T-\log T_0)^2/(2\sigma ^2))
\end{equation} 
with a peak temperature of $T_0=5$ MK, width $\sigma=0.1$ and $\xi_0=10^{36}$ $\mathrm{cm^{-3}K^{-1}}$ where $\xi_0$ was chosen such that the total emission measure is one order of magnitude larger than in the first case to simplify the presentation in Figure \ref{fig:synDEM}.
We define the total emission measure as 
\begin{equation} \label{eq:totem}
EM=\int_{T_{min}}^{T_{max}} \xi(T)dT
\end{equation} 
where $T_{min}=0.01$ MK and $T_{max}=100$ MK in the synthetic case. We again compare the electron flux spectrum inferred from the DEM with a single temperature Maxwellian. Since a Gaussian DEM function does not represent a single temperature distribution some discrepancy, especially at higher energies, can be expected. As Figure \ref{fig:synDEM} demonstrates the inferred mean electron flux spectrum is still in good agreement with the single temperature model at low energies, but diverges at higher energies, due to the contribution of non-zero DEM at temperatures higher than the peak temperature that was used for the Maxwellian model. This interpretation is supported by the fact that adding an additional Maxwellian temperature model with temperature 9 MK and emission measure 0.09 times the total EM of the first DEM model adequately describes the high energy component and leads to a better agreement except for the highest temperatures (see Figure \ref{fig:synDEM}, blue dashed lines).
\section{DISTRIBUTION FUNCTIONS FROM RHESSI AND SDO/AIA OBSERVATIONS}
We now apply the method to two events for which there were good simultaneous RHESSI and AIA observations. The first event was a GOES class C4.1 event on August 14th 2010 (henceforth SOL2010-08-04T10:05) with HXR peak time around 09:46 UT. It displays a EUV loop that is co-spatial with a SXR loop seen in RHESSI images. Because the event was relatively weak, the regions of interest were not saturated in all AIA wavelength channels \citep[cf.][]{2012ApJ...760..142B}. The second event was a GOES M7.7 limb event on July 19th 2012 (henceforth SOL2012-07-19T05:58) with the first HXR peak at around 05:22 UT, showing two footpoints, a SXR coronal source, and a HXR above-the-looptop source. In this event the exposure times in all AIA wavelengths were short enough to provide unsaturated images even during the flare-peak. For both events we chose the time interval with highest observed X-ray energies within the limitations given by both instruments such as saturation of AIA images and pile-up in RHESSI spectra. For both events the mean electron distribution function at energies below $\sim$ 1 keV is calculated using AIA DEM measurements. RHESSI observations are used to find the distribution function above 3 keV up to the highest observed X-ray energies ($\sim$ 25 keV for SOL2010-08-04T10:05 and 80 keV for SOL2012-07-19T05:58) .
\subsection{SOL2010-08-14T10:05}
For this flare we chose a time-interval during the rise phase of the flare when AIA images were not saturated and a small non-thermal tail was observed with RHESSI. Panel a) in Figure \ref{fig:aug14analysis} shows a 131 \AA\ AIA image overlaid with 20 \%, 30 \%, and 50 \% contours from a RHESSI 8-10 keV CLEAN image, calculated using grids 3 - 8 and natural weighting with clean-beam-width-factor 1.7 (corresponding to an effective beam FWHM of 9.1 arcsec).
The RHESSI source is co-spatial with the top of the EUV loop and extending downward along the legs of the EUV loop. The bulk of the EUV emission originates from ribbons at the bottom of the loop and AIA images in these regions consequently are saturated in most wavelength bands. Here we focus on the part of the loop that is co-spatial with the RHESSI source where no saturation occurred and the regularized inversion to find the DEM \citep{2005SoPh..226..317K} could be made with high confidence.
\subsection*{RHESSI spectrum and electron flux distribution}
The RHESSI full Sun spectrum suggests the presence of a thermal component and a small non-thermal tail extending to about 30 keV. The spectrum was fitted between 09:42 and 09:43 UT (attenuator state 0), during the rise phase of the flare but before pile-up started to dominate energies above 10 keV \citep{Smith02}, with a single-temperature component of 10.5 MK temperature and emission measure of $3\times 10^{47}$ $\mathrm{cm^{-3}}$, and a thin-target power-law component with spectral index $\delta = 3.9$ and low-energy cut-off of 10 keV. In addition to the RHESSI temperature, the temperature and emission measure from GOES was determined for the same time-interval, giving $T_{GOES}$=9.5 MK and $EM_{GOES}=5\times 10^{47}$ cm$^{-3}$. RHESSI images suggest that there is faint footpoint emission above $\sim$ 25 keV, but the bulk of the non-thermal emission below this energy seems to originate from the flaring loop. The mean electron flux spectrum for the thermal component is then just the Maxwellian distribution (Equation (\ref{eq:totmaxwell})). The electron flux spectrum for the non-thermal power-law component is easily found under the thin-target assumption \citep[e.g.][]{Ta88}. 
\subsection*{Mean electron flux spectrum from AIA differential emission measure}
\begin{figure*}
\begin{center}
\includegraphics[height=15cm]{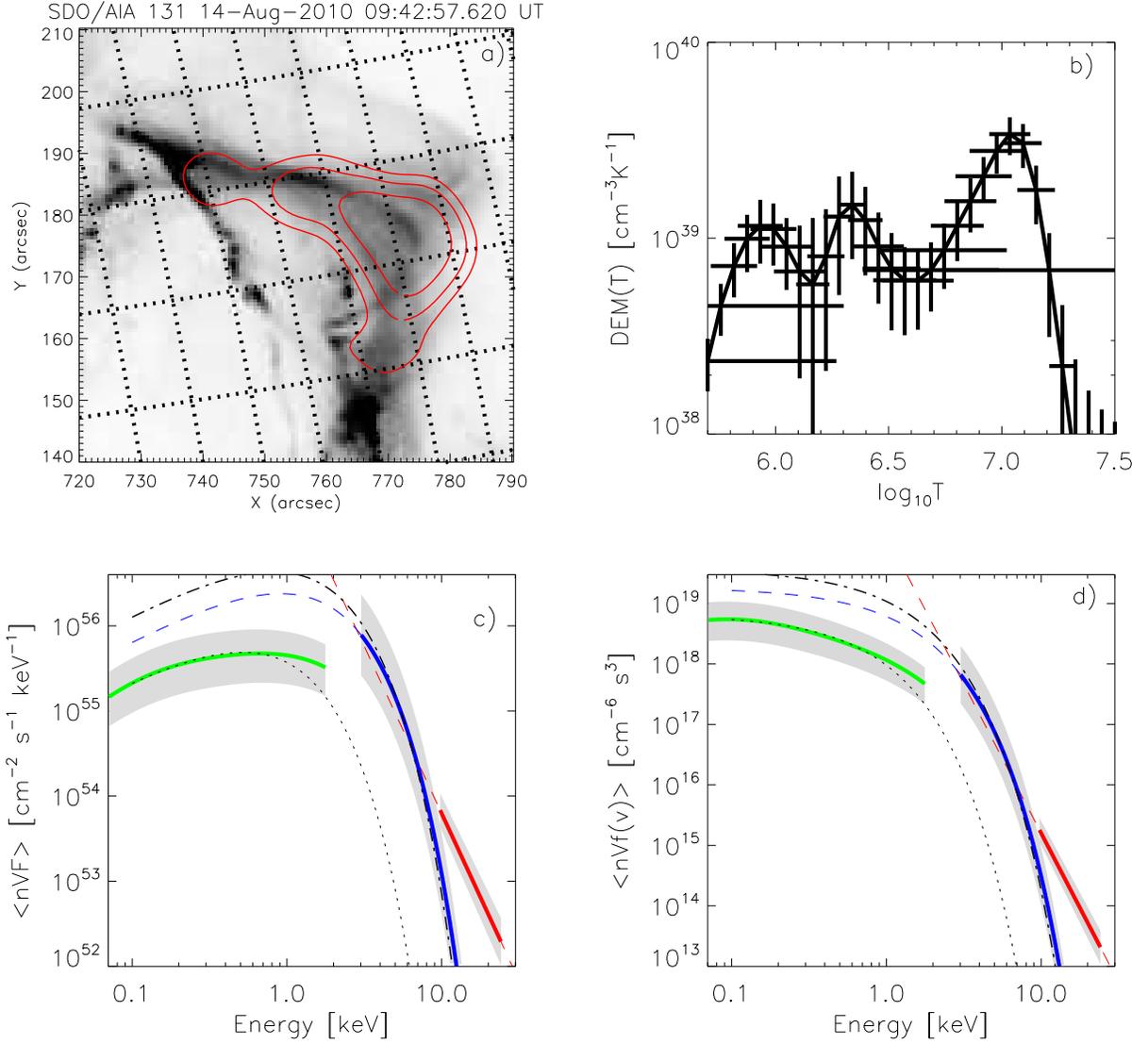}
\end{center}
\caption {Panel a: AIA 131 \AA\ image overlaid with RHESSI contours (red; 20, 30, 50 \% in 8-10 keV CLEAN image). Panel b: AIA DEM from area corresponding to RHESSI 50\% contours in 8-10 keV CLEAN image. Panel c: mean electron flux spectrum derived from AIA (green) and RHESSI thermal fit (blue) and non-thermal fit (red). The gray shaded area gives the confidence interval. Dashed lines indicate the extension of the flux to energies that were not observed with the respective instrument. Dash-dotted line: Electron flux spectrum from GOES temperature and emission measure. The dotted line represents a Maxwellian distribution with $T=6$ MK and $EM=4.5\times 10^{46}$ $\mathrm{cm^{-3}}$ for illustration (not from an actual fit). Panel d: mean electron velocity distribution.}
\label{fig:aug14analysis}
\end{figure*}

The AIA DEM per area $\mathrm{(cm^{-5}K^{-1})}$ was calculated using the regularized inversion method developed for RHESSI by e.g. \citet{2005SoPh..226..317K} and adapted for SDO/AIA by \citet{2012A&A...539A.146H}. Such DEM can be calculated pixel by pixel or for any finite area \citep[e.g.][]{2013A&A...553A..10H,2012ApJ...760..142B}. For direct comparison with RHESSI results the EUV region that is co-spatial with the RHESSI emission was analyzed under the assumption that the same emitting plasma is observed in all wavelengths. RHESSI full Sun spectra are dominated by the flaring emission and it is often assumed that the bulk of the emission originates from a region that corresponds to the size of the 50\% contours in a RHESSI image. However, AIA images clearly outline the whole loop, as do the RHESSI contours down to 20\%. We therefore calculate the total AIA DEM from several areas, namely the ones corresponding to 50\%, 30\%, and 20\% contours in the RHESSI 8-10 keV CLEAN image and use the result from the 50 \% and the 20 \% contours as confidence interval. 
The DEM from within the 50\% contour is shown in panel b) of Figure~\ref{fig:aug14analysis}. The error bars represent the uncertainties of both, the DEM and the temperature, i.e. the effective temperature resolution, obtained from the regularized inversion \citep[see][for full details]{2012A&A...539A.146H}. The DEM suggests the presence of two main temperature components, a weak one at 2 MK and one at 10 MK. The low temperature component can most likely be attributed to background emission while the high temperature component is dominated by flaring emission \citep[see also][]{2012ApJ...760..142B}. Note that in this case we did not impose a positivity constraint on the reconstructed DEM (see Appendix) because the assumption of a positive DEM is quite strong and only correct in the case of purely thermal plasma. From the DEM, the mean electron flux spectrum is calculated using the method described in Section~\ref{sec:theory}. Figure \ref{fig:aug14analysis} shows the mean electron flux spectrum in units of $\mathrm{(electrons\,cm^{-2}keV^{-1}s^{-1})}$ as a function of energy from the combined AIA and RHESSI observations, where we use the result from the 50 \% contours and 20\% contours as confidence interval. Dividing by energy and multiplying with $m_e^2$ we can also display the spectrum as a velocity distribution function $\langle nVf(v) \rangle$ (Figure~\ref{fig:aug14analysis}, panel b)). The distribution found from AIA is consistent with a Maxwellian of temperature $T=6$ MK and emission measure $EM=4.5\times 10^{46}$ $\mathrm{cm^{-3}}$, but deviates from the Maxwellian distribution at energies greater than 1 keV. The extrapolation of the RHESSI thermal distribution into the AIA regime is a factor $\sim 3$ larger than the distribution from AIA. We discuss several reasons for this discrepancy in Section~\ref{sec:dandconc}. The overall distribution over all energies resembles particle distributions often found in the solar wind with a core-halo-strahl structure \citep[see][for a review]{2006LRSP....3....1M}.
\subsection{SOL2012-07-19T05:58}
\begin{figure*}
\begin{center}
\includegraphics[height=13cm]{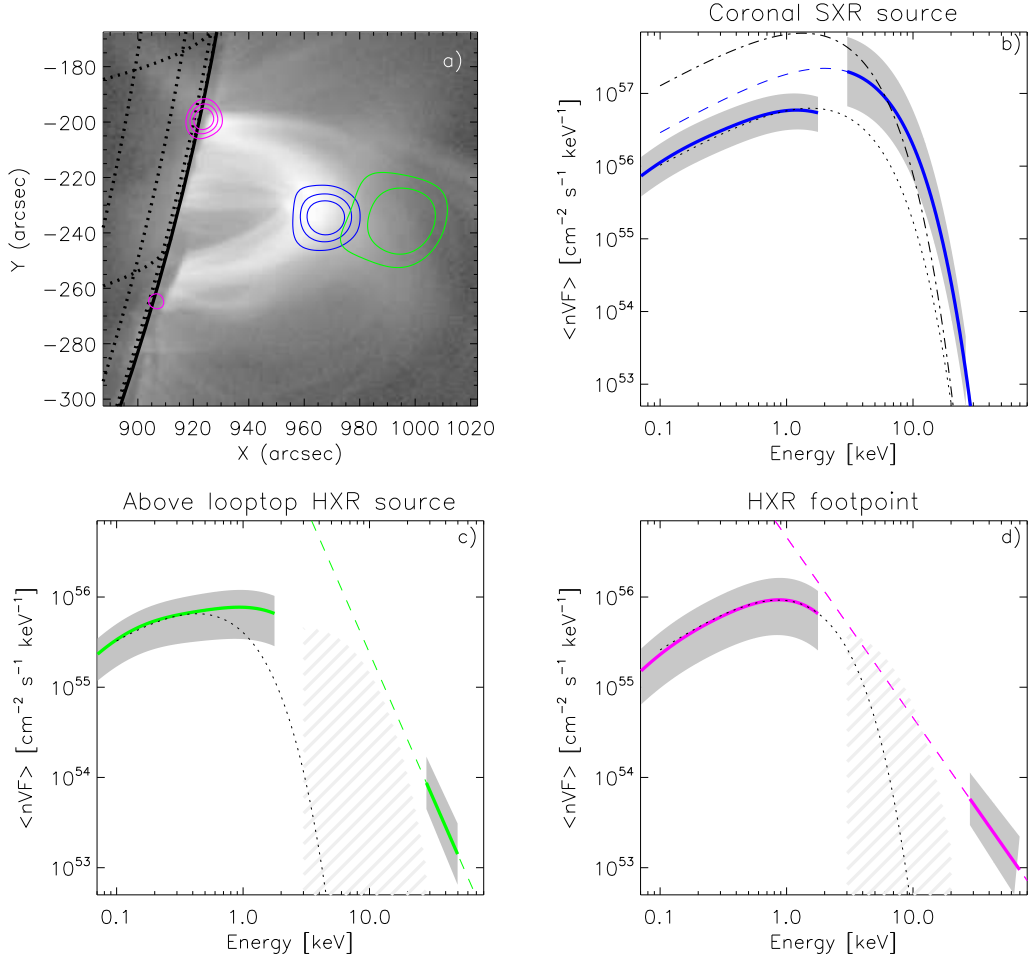}
\end{center}
\caption{Panel a: AIA 131 \AA\ image of SOL2012-07-19T05:58 overlaid with RHESSI contours indicating the HXR footpoint (purple), the coronal SXR source (blue), and the HXR above-the-looptop source (green). Panel b: Mean electron flux spectrum derived from AIA DEM measurement and RHESSI spectral fitting in coronal SXR source. The dotted line indicates a Maxwellian distribution with $T=18$ MK / $EM=10^{48}$ cm$^{-3}$. Panel c: same as panel b) but for the above-the-looptop source. The dotted line gives a Maxwellian with $T=5$ MK / $EM=5.5\times 10^{46}$ cm$^{-3}$. The upper edge of the shaded gray area at intermediate energies represents the upper limit of the flux found from a thermal fit to the reconstruction noise of the RHESSI image (see Section \ref{rr}). Panel d: same as panel c) but for northern footpoint. The dotted line gives a Maxwellian with $T=10$ MK / $EM=10^{47}$ cm$^{-3}$ (not from an actual fit).}
\label{fig:julymeanelectronflux}
\end{figure*}

For this limb event three distinct sources were observed with RHESSI (SXR coronal source, HXR above-the-looptop source, HXR footpoints, see Figure \ref{fig:julymeanelectronflux}. The event has been analyzed in detail by \citet{2013ApJ...767..168L} with respect to several aspects of its time evolution and with a focus on the coronal densities by \citet{Kr13}. AIA exposure times where as short as 0.2 seconds during the course of the flare. Thus there are unsaturated images in all wavelength channels even at the flare peak-time. Here we focus on the same time-interval (05:20:30 to 05:23:02 UT, attenuator state 1) used by \citet{Kr13} who analyzed the first HXR peak using imaging spectroscopy, and we present mean electron distribution functions for three different sources observed by RHESSI: the SXR coronal source, the HXR above-the-looptop source, the northern footpoint. A weak second footpoint that was likely occulted was also observed. STEREO images of the region suggest the presence of loops or a loop-arcade for which, as seen from Earth, the southern footpoint would be occulted. Note that the northern footpoint was also likely partly occulted. Footpoint sources at higher energies are formed deeper down in the chromosphere and have a vertical extent of at least one Mm \citep{Ba11a,Koet10}. Thus the higher energies will be more occulted relative to lower energies resulting in a softer HXR spectrum and lower observed flux. Therefore the inferred electron flux spectrum from an occulted footpoint will represent a lower limit.
\subsection*{RHESSI spectrum and electron flux distribution} \label{rr}
Using full Sun spectroscopy for the SXR coronal source and fitting a single temperature thermal model \citet{Kr13} found $T=23$ MK, $EM=4\times 10^{48}$ $\mathrm{cm^{-3}}$. The GOES temperature and emission measure during the corresponding time interval were $T_{GOES}=15.5$ MK, $EM_{GOES}=10^{49}$ cm$^{-3}$.
Imaging spectroscopy of the above-the-looptop HXR source resulted in a electron spectral index $\delta=3.2 \pm 0.2$ between 28 - 50 keV, the spectral index of the northern footpoint was $\delta=2.0 \pm 0.2$ between 20 - 70 keV, where a thin-target assumption was used in both cases. In the intermediate energy-range between $\sim$ 14 to 20 keV the coronal source and the footpoints are seen in the images. Such a complex source structure makes image reconstruction with RHESSI rather difficult and it is often problematic to distinguish reconstruction noise from actual source emission \citep{Hur02,Ba06}. We therefore only estimated an upper limit of the electron flux in this energy-range by fitting a thermal (Maxwellian) component to the X-ray spectrum (compare Figure \ref{fig:julymeanelectronflux}). 
\subsection*{Mean electron flux spectrum from AIA differential emission measure}
The AIA differential emission measure and subsequently the mean electron flux spectrum was calculated within the three regions as defined by contours from a RHESSI CLEAN image at 5-7 keV using grids 3,5,6 with uniform weighting (giving an effective CLEAN beam FWHM of 8.7 arcsec) in the case of the coronal source. For the HXR sources two-step CLEAN \citep{2011ApJ...742...82K} images at 30-70 keV were made using grids 1-5 (uniform weighting, effective beam FWHM=3 arcsec) to image the footpoints and grids 4-9 (uniform weighting, effective beam FWHM=17 arcsec) to image the above-the-looptop source. For the HXR footpoints and SXR coronal source the 30 \%, 50 \% and 70 \% contours were used, and we use the 30 \% and 70 \% contours as a confidence interval. The coronal HXR source showed extended emission over a large area and we use the 50 \% and 70 \% contours to give a confidence interval, since the 30 \% contours overlap with the coronal SXR source.

\begin{figure}
\begin{center}
\includegraphics[height=10cm]{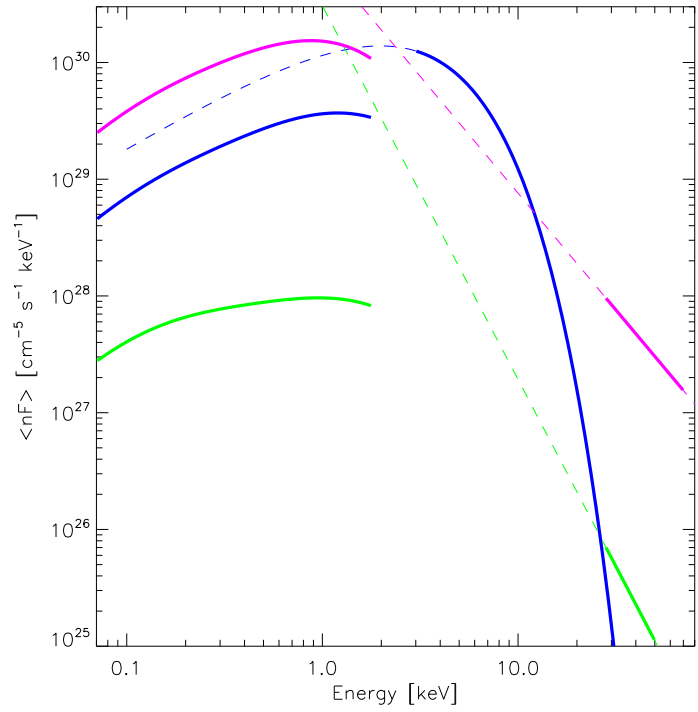}
\end{center}
\caption {Comparison of mean electron flux spectrum \textit{per unit volume} between the different sources of SOL2012-07-19T05:58 (compare Figure \ref{fig:julymeanelectronflux}). Blue lines: SXR coronal source. Purple: HXR footpoints. Green: HXR above-the-looptop source. }
\label{fig:julymeanelectronfluxpervolume}
\end{figure}

We find that the mean electron flux spectrum derived from the AIA data for the coronal SXR source at energies between 0.1 - 2 keV can be approximated with a Maxwellian distribution of temperature 18 MK and emission measure  $10^{48}$ $\mathrm{cm^{-3}}$. As in the previous event, the extrapolation of the RHESSI spectrum to energies below 3 keV is a factor of $\sim$ 2 larger than the AIA values. The electron flux spectrum derived from AIA for the HXR footpoint is consistent with a Maxwellian at temperature $T=10$ MK and emission measure  $EM=10^{47}$ $\mathrm{cm^{-3}}$. The spectrum of the HXR above-the-looptop source is consistent with a Maxwellian at temperature 5 MK and emission measure $5.5\times 10^{46}$ $\mathrm{cm^{-3}}$ only up to about 0.6 keV and deviates from the Maxwellian distribution at higher energies. For a direct comparison of the number of electrons between the distinct regions it is useful to compare the mean electron spectra per unit volume. We define the volume as $V=A^{3/2}$ where $A$ is the area of the 50 \% RHESSI contours of the SXR coronal source and the HXR above-the-looptop source. For the footpoint the effective area over which EUV emission is observed is smaller than the 50\% contour by about a factor 3 and the EUV emission originates from a height above the centroid of the HXR emission (Figure \ref{fig:julymeanelectronflux}, panel a)). Thus we define the area of EUV emission from the footpoint region as the area of the 50\% HXR contour divided by 3. Figure \ref{fig:julymeanelectronfluxpervolume} shows a comparison between the mean electron flux spectra per unit volume from each of the sources. The figure indicates that the number of electrons per unit volume is highest in the footpoint (a factor of 6 larger than the SXR coronal source) and is lowest in the HXR above-the-looptop source (one order of magnitude smaller than in the SXR coronal source). 
\section{DISCUSSION} \label{sec:dandconc}
We present and demonstrate a new method to infer the mean electron flux spectrum from differential emission measure (DEM) results derived from AIA images in different EUV wavelength channels. The method was applied to two well observed flares. The first event displays a EUV loop that is co-spatial with a SXR loop observed by RHESSI, and the mean electron flux spectrum was derived for the loop as a whole. The second event can be divided into HXR footpoints, SXR coronal source, and HXR above-the-looptop source. Each of the three sources was analyzed separately. In all presented cases the result from AIA observations is a factor of 2.5-8 lower than the extrapolated thermal model from the RHESSI fit. There are several possible explanations for this discrepancy: the RHESSI fit, insensitivity of AIA to high temperatures, the method to infer the DEM, and two physically distinct particle populations. The RHESSI data-analysis tools are developed to a high standard and it is believed that the instrument is well understood. In combination with good counting statistics in flares this generally leads to rather small uncertainties of the fitted parameters \citep[e.g.][]{2013ApJ...769...89I}. However, for weak flares and when using imaging spectroscopy it is often possible to fit different models (i.e. different temperature and emission measure) with equal $\chi^2$ value. Further, the attenuator state affects the low-energy limit to which a confident fit can be made \citep{Smith02}. This it not an issue for SOL2010-08-14T10:05, since no attenuator was in place, but with the thin attenuator in place during SOL2012-07-19T05:58 the spectral fit could only be performed down to 6 keV. In addition it has recently been found that the thermal blanket thickness of the RHESSI instrument could be overestimated by up to 30 \% (Brian Dennis, private communication). This affects thermal fits in attenuator state 0 and in the case of SOL2010-08-14T10:05 could explain the discrepancy at least partly. All of these factors introduce an additional uncertainty not reflected in the purely statistical errors. We therefore assume an upper limit on the uncertainty of the fitted emission measure of a factor of 2. This could bring the observed RHESSI emission measure down within the range of the AIA value. Comparison with GOES gives a slightly higher emission measure ($5\times 10^{47}$ cm$^{-3}$ opposed to $3\times 10^{47}$ cm$^{-3}$) and slightly lower temperature ($9.5$ MK opposed to $10.5$ MK) for SOL2010-08-14T10:05. The difference is more striking for SOL2012-07-19T05:58 ($10^{49}$ cm$^{-3}$ opposed to $4\times 10^{48}$ cm$^{-3}$ and $15.5$ MK opposed to $23$ MK). RHESSI emission measures tending to be smaller than those derived from GOES observations is a common pattern observed repeatedly in the past \citep[eg.][]{Ha08, Ba05}. One explanation is that this reflects RHESSI's limited sensitivity to temperatures below $\sim$ 8 MK, temperatures to which GOES is sensitive. However, it has to be noted that GOES observations are full Sun measurements with no way of knowing the exact position and extent of the source. In Figures \ref{fig:aug14analysis} and \ref{fig:julymeanelectronflux} we assumed the same area as measured with RHESSI but this is only an approximation. In summary the fitted RHESSI emission measure in cooler flares is probably a lower limit of the true emission because of reduced sensitivity, on the other hand the emission measure could be over-estimated in the attenuator 0 state. Another explanation in certain flares is the main temperature sensitivity range of AIA. The response function of the high-temperature wavelength channels 131 \AA\ and 193 \AA\ peaks at 12 MK and 16 MK, respectively and falls off sharply at higher temperatures. If the bulk of the plasma is at higher temperatures than 12-16 MK, AIA will not be sensitive to its signatures, resulting in an underestimated total emission measure. This could explain the discrepancy in the SXR coronal source of SOL2012-07-19T05:58, for which a RHESSI temperature of 23 MK was found. On the other hand the RHESSI fit of SOL2010-08-14T10:05 suggests a temperature of 10.5 MK which falls within the main AIA sensitivity range. To investigate the systematics of the DEM code and the influence of parameters such as the data-error and use of positivity constraint in the regularized inversion we performed a systematic study using idealized Gaussian DEMs with different peak temperatures. From these, the expected AIA data-number (DN) was calculated and used to find the DEM from regularized inversion (see Appendix). In addition, a model with two Gaussian DEMs was used, emulating the often found double-peak structure of the DEM. Figure~\ref{fig:appendix1} shows the model DEM and the reconstructed DEM for several values of peak temperature and several relative intensities in the case of two Gaussians. Generally there is a better agreement between model and reconstructed DEM without the imposed positivity constraint. In the case of purely Poisson error without any systematic terms the shape of the DEM is well reconstructed except for peak temperatures higher than $\log T\sim 7$ where some of the DEM seems to be ``redistributed'' to low temperatures at around $\log T=6.2$. The same but opposite effect occurs for very low peak temperatures of around $\log T\sim 6$. This is probably due to the 193 \AA\ response which has two peaks, one at $\log T=6.2$ and one at $\log T=7.2$. This effect is worsened when the positivity constraint is applied. In all cases, the total emission measure defined as $EM=\int_{\log T=5.7}^{\log T=7.5} \xi(T) dT$ is reduced relative to the total model emission measure (Figure~\ref{fig:appendix2}) by a factor of up to 30 when the positivity constraint is used and up to 2 without positivity constraint. However, the effect is not systematic enough to allow for implementation of an empirical correction in the DEM reconstruction.

There is also the possibility of several distinct particle populations with different temperatures being present. The mean electron flux spectra derived from AIA at the footpoints and the HXR above-the-looptop source in SOL2012-7-19T05:58 are consistent with a Maxwellian distribution, but with a deviation near 1 keV while RHESSI observes a power-law distribution at energies above 25 keV. Unfortunately there were no reliable RHESSI observations possible between 3 to $\sim$ 25 keV for these two sources. We fitted a thermal component to the spectrum from imaging spectroscopy of the two sources in this energy range. While potentially not physically meaningful because the spectrum mainly consist of reconstruction noise from the different sources this provides a reasonable upper limit of the flux (compare Figure \ref{fig:julymeanelectronflux}) and suggests that the presence of a second Maxwellian distribution similar to the one found in the SXR coronal source and in the loop of SOL20100814T10:05 is feasible.
\section{CONCLUSIONS}
Combining X-ray observations with EUV observations from RHESSI allows the mean electron flux spectrum from different sources in solar flares to be inferred from 0.1 keV up to several tens of keV, thus enabling diagnostics of the low energy component of the spectrum inaccessible with RHESSI alone. There is still a gap in spectral coverage where uncertainties are rather large at the intermediate energies between about 1 - 3 keV where neither instrument is sensitive for reliable measurements. This is the most likely cause for the discrepancy between the total emission measures of RHESSI and AIA.
The total spectrum can be described as the combination of a Maxwellian core, a secondary ``halo''-component and a non-thermal tail similar to distributions often seen in the solar wind \citep[e.g.][]{1997AdSpR..20..645L}. An analytical description of the total solar flare spectrum and a quantitative comparison with solar wind spectra will be the subject of future work.
%
\acknowledgments
Financial support by the European Commission through the FP7 HESPE network (FP7-2010-SPACE-263086) is gratefully acknowledged. This work is supported by an STFC Consolidated Grant (E.P.K.) and by the Swiss National Science Foundation (M.B.). We thank the referee for valuable comments and suggestions.
\appendix
\section{SIMULATED DIFFERENTIAL EMISSION MEASURES} \label{s:appendix}
\begin{figure*}
\begin{center}
\includegraphics[width=16cm, height=16cm]{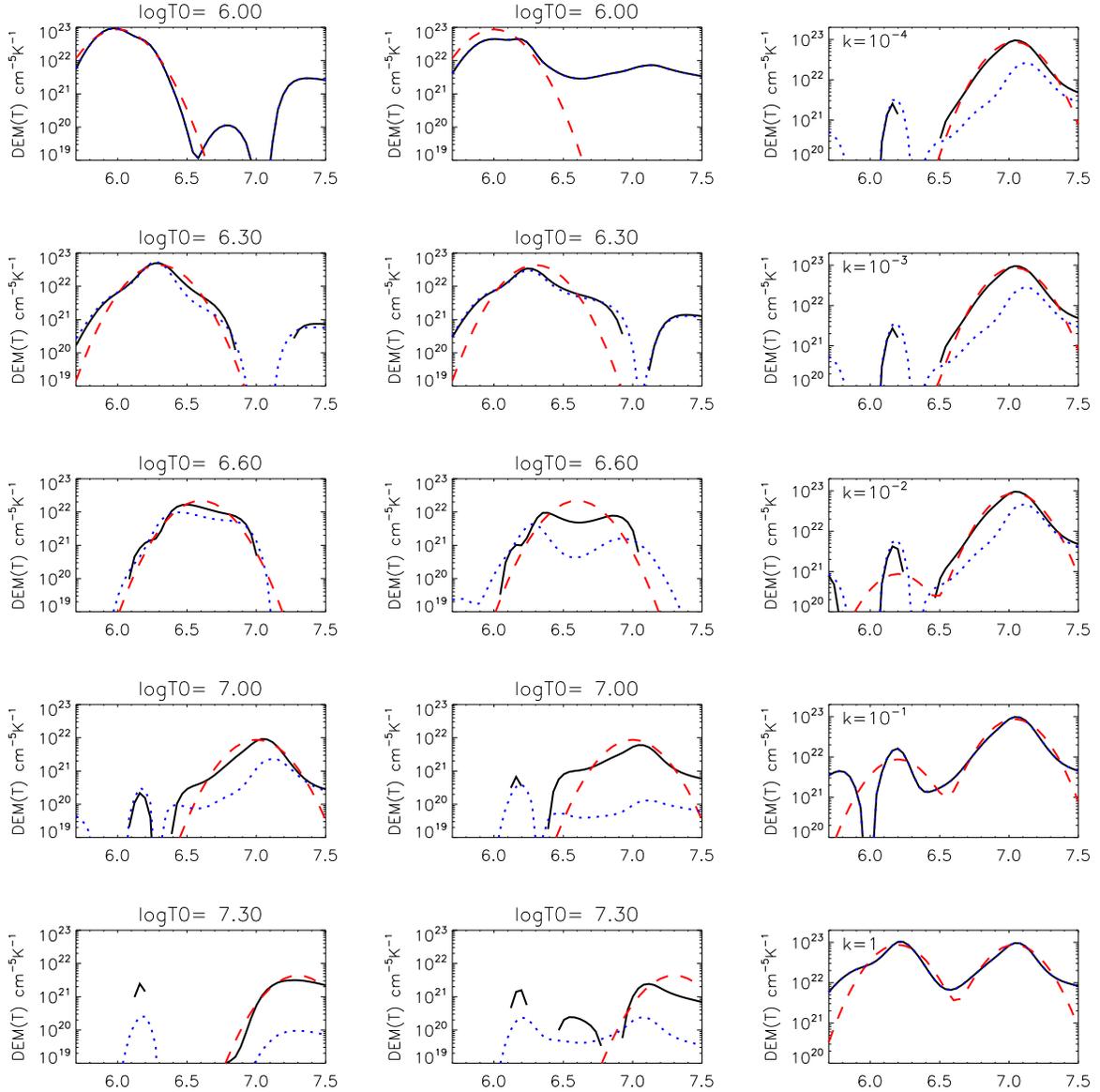}
\end{center}
\caption {Left (top to bottom): Idealized Gaussian DEMs for a number of peak-temperatures $\log T_0$ with an assumed error of the AIA data-number (DN) of $DN_{err}=\sqrt{DN}$. Middle: same as left column but with an included systematic error: $DN_{err}=\sqrt{DN+(0.2\times DN)^2}$. Right: reconstructed DEM from two Gaussian components. The dashed red lines give the input model DEM, the black solid lines are the reconstructed DEM without positivity constraint, the dotted blue lines are the reconstructed DEM with positivity constraint. The factor k gives the ratio between the DEM of the cold component and hot component. }
\label{fig:appendix1}
\end{figure*}

\begin{figure*}
\begin{center}
\includegraphics[height=16cm]{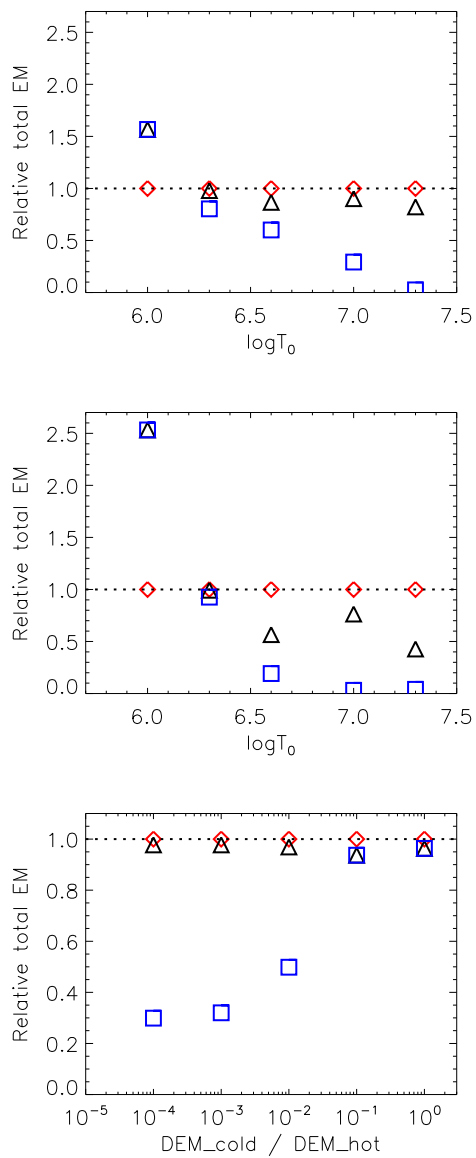}
\end{center}
\caption {Top: Total EM relative to total model EM as a function of $\log T_0$ for data-number error of $DN_{err}=\sqrt{DN}$. Middle: Same as top but with data-number error of $DN_{err}=\sqrt{DN+(0.2\times DN)^2}$. Bottom: Total EM relative to total model EM in the case of a double-Gaussian model as a function of relative peak emission measure. Red diamonds: model DEM. Black triangles: no positivity constraint. Blue squares: with positivity constraint. }
\label{fig:appendix2}
\end{figure*}

To investigate the influence of parameters such as the data-error and use of positivity constraint in the regularized inversion method that was used for the analysis of the presented events we performed a systematic study using idealized Gaussian DEMs. An idealized Gaussian DEM in logT of the form:  $\frac{DEM_0}{\sqrt{2\pi \sigma}}\exp[-(\log T-\log T_0)^2/(2\sigma^2)]\times 10^6/T_0$ was used where $\sigma=0.15$ for 6 values of $\log T_0$ between 5.8 and 7.5 and $DEM_0=3.27\times 10^{22}$ $\mathrm{cm^{-5}K^{-1}}$. From this, the expected AIA data-number (DN) was calculated and used to find the DEM from inversion for an assumed data error of $DN_{err}=\sqrt{DN}$ and including a ``systematic error'' $DN_{err}=\sqrt{DN+(0.2\times DN)^2}$ to account for calibration uncertainties \citep[e.g.][]{2010ApJ...714..636L}. In both cases, the DEM was found once without other constraints and once imposing a constraint for positive DEM at all temperatures. In a second step, a DEM with two Gaussian components (with peak-temperatures $\log T_{cold}=6.2$ and $\log T_{hot}=7.04$, where $DEM(\log T_{cold})=k*DEM(\log T_{hot})$ with k between $10^{-5}$ and $1$) was used to emulate the often observed two-peak structure. Figure~\ref{fig:appendix1} shows the input model along with the reconstructed DEMs for all temperatures. We can now define the total emission measure as the integral over the DEM $EM=\int_{\log T=5.7}^{\log T=7.5} \xi(T) dT$ and compare this as a function of $\log T_0$ and k (Figure \ref{fig:appendix2}).

%
\bibliographystyle{apj}
\bibliography{mybib,refs_rhessi}

\end{document}